\documentclass[]{interact}

\usepackage{comment}
\usepackage{graphicx}
\usepackage{url}
\usepackage[x11names]{xcolor}
\usepackage{framed}
\usepackage{quoting}
 \colorlet{shadecolor}{LavenderBlush2}
\newenvironment{shadedquotation}
{\begin{shaded*}
}
{\endquoting
\end{shaded*}
}
\usepackage{hyperref}

\usepackage{epstopdf}
\usepackage[caption=false]{subfig}

\usepackage[natbibapa,nodoi]{apacite}
\begin{document}


\title{Designing Child-Centered Content Exposure and Moderation}

\author{
\name{Bel\'en Sald\'ias\thanks{CONTACT Email: belen@mit.edu}}
\affil{Massachusetts Institute of Technology, USA}
%
}

\maketitle

\begin{abstract}
Research on children's online experience and computer interaction often overlooks the relationship children have with hidden algorithms that control the content they encounter. Furthermore, it is not only about how children interact with targeted content but also how their development and agency are largely affected by these. By engaging with the body of literature at the intersection of i) human-centered design approaches, ii) exclusion and discrimination in A.I., iii) privacy, transparency, and accountability, and iv) children's online citizenship, this article dives into the question of ``How can we approach the design of a child-centered moderation process to (1) include aspects that families value for their children and (2) provide explanations for content appropriateness and removal so that we can scale (according to systems and human needs) the moderation process assisted by A.I.?''.

This article contributes a sociotechnical highlight of core challenges and opportunities of designing child-centered content control tools. The article concludes by grounding and characterizing design considerations for a child-centered, family-guided moderation system. We hope this work serves as a stepping stone for designers and researchers pursuing children's safety online with an eye on hidden agents controlling children's online experiences and, by extension, the values and opportunities children are exposed to.
\end{abstract}


\section{Introduction}
Data and moderation policies vary across platforms, making it difficult to hold designers, architects, and developers of technology accountable to their policies or holding them to standards across different technologies~\citep{nissenbaum1996accountability, eubanks2014want, barassi2019datafied}, and even when available, these policies and moderation guidelines are typically written using concepts that tend to be ambiguous, unclear or lacking transparency for the majority of the target users, leading them through resignation to accept these terms~\citep{draper2019corporate}. While one can argue that we are free to opt-out from using certain technologies, the reality is that more and more we are being coerced into using them, and we need to find an actionable path ahead~\citep{barassi2019datafied}.

Before algorithmic personalization~\citep{Netflix-personalization, google-personalization}, when individuals were initiated into the online world and learned about their identity and possibilities online, most often their online interactions were influenced by other online human beings, many times unidentified or with misleading intentions, but human nonetheless. Technology and media seemed more like an exploration-targeted medium where intelligent engagement entailed social connection,  as opposed to an exploitation-targeted medium where, as of today, social connection is seen as a means to engage users and capitalize on their engagement~\citep{o2016weapons}; treating people as targetable users, disregarding their everyday values, autonomy, self-determination, and intentions by adopting a reductive form of digital citizenship~\citep{third2016rethinking}. Today the online space is crowded with ``intelligent'' agents or machine learning agents, which are granted the ability to profile us, shape, and bound our choices at large scale.

Today it is not only about how children interact with targeted content, algorithms or automated agents, but also how their development and agency are affected and conditioned by these interactions. Because of children's developmental age, algorithms are increasingly shaping their lives.

Fortunately,  in the early days of the internet, exposure to it as a child had no strings attached for that child's future (a.k.a. current) self, even though they may had crossed some age-appropriateness barriers.

Online children safety is not only about data privacy, counteracting malicious human agents, or/and preventing bullying. There are also pressing risks and challenges in children interacting with ``intelligent'' agents (through content) that reflect values that are not necessarily those that children or their families and communities wish for them as they are growing and developing~\citep{bridle2017something}.

The need for robust, socially-sensitive, and in particular children-centered, artificial intelligence (A.I.) technologies is becoming more pressing. Our interest in pursuing this area started by questioning how we could make the internet a safer place for youth, and this work brings a perspective to the ways in which children are unintentionally or intentionally being exposed to content online---focused on that content decided by A.I. algorithms---and ways in which we can face primary ethical considerations from a design+A.I. perspective, specifically as technology is being interacted with and shaping our children~\citep{kids-shapeable}.

\section{Children exposure to online content}
\label{sec:content-exposure}

Social media platforms, search engines, and web navigation, in general today, rely on multiple methods to engage users in accessing and interacting with the information they contain. This engagement can be highly beneficial in exposing children to a variety of well-rounded learning opportunities and connecting them to resources with unique facets only available through the internet (e.g., remote friendship and spaces for identity exploration)~\citep{livingstone2017maximizing}. Nevertheless, it is crucial to notice that a system could push to cocoon children into specific optimization criteria defined by external agents or policies by exploiting personalized recommendations~\citep{pariser2011filter, van2020teenagers}.

To better understand the extent to which some of these content exposure and engagement mechanisms are everywhere in today's online experience, it is essential to look at how these surfaced and fomented. In the early 2000s, when Netflix intended to grow its library to 100,000 titles open for all-you-can-eat consumption, they looked for a strategy to help their clients find movies more efficiently, which included search and recommendations~\citep{Netflix-personalization}. These two features allowed Netflix to start collecting a surplus of data such that they could predict with high confidence those movies that your friends may enjoy or that you may enjoy based on your viewing history, movie ratings, demographic information, and your friends' collected data. The main aim was to collect data to train algorithms to connect clients to movies they would enjoy. At the same time, Google Search was starting what is today the primary handler of online search requests, and whose revenue comes almost all from personalized ads (Google Ads) prompted from users' search requests~\citep{google-history, google-personalization}. Google's monetization of data captured through monitoring people's movements and behaviors online and in the physical world has had more profound repercussions than what anyone had anticipated then~\citep{zuboff2019age, google-pregnant}. Children are not the exception, and while content moderation tools have been in place for a long time, children's exposure to online content deserves attention; children are subject to not only being monitored but also influenced and shaped by these personalization systems (some examples presented by~\citet{bridle2017something, tan2018s}).

The unique focus of this work looks at the diverse shapes in which children get exposed to content online today, varying on the level of endedness (open-ended vs. closed-ended) and personalization (generic vs. targeted vs. restricted). Of course, these are not unique to children, but they bring particular challenges when interacting with them.

\subsection{Stream of content (open-ended exposure)}
\label{subsec:open-ended-exposure}
Whenever we spend time online, we have access to a stream of content. When it comes to children, open-ended exposure allows them to discover topics and opportunities that they may not have envisioned for themselves before~\citep{coleman2007adolescence, livingstone2017maximizing, chiong2014pioneering}. For example, learning about what others are doing, possible career paths, seeing Ads about educational tools, and playing video games. While all these scenarios can be highly beneficial to children, there are still some technical and ethical challenges to be addressed to enjoy these goodnesses fully:

\begin{itemize}
    \item Hard to converge: This excess of information can lead to high exposure to irrelevant information with a lack of depth in specific knowledge in children pursuing open-ended media engagement. To address this challenge, technology platforms now offer, almost by default, search engines or filtering software capabilities.

    \item Adequacy for minors: Even when children navigate the internet through a browser with a parental-control system activated, they are likely to encounter different kinds of inappropriate content. To comply with COPPA and GDPR~\citep{coppa-gdpr}, social media platforms are set to discourage children from using their services. However, because these measures do not actively prohibit or ban minors from accessing the services--since it is up to the user to disclaim their age--, children are still an active part of these communities. 
\end{itemize}

\subsection{Search engines (closed-ended exposure)}
\label{subsec:closed-ended-exposure}
Directed search and navigation is a very effective mechanism to converge to the content we are looking for. Search engines and features that allow us to filter content have brought enormous advantages and efficiencies to our online experience, allowing us to categorize those areas of most interest for us. However, there are some challenges that we as adults encounter and need to learn how to deal with---for example representational biases and how these can affect children self-image if validated or discriminated at large-scale by an A.I. system / search engine~\citep{noble2013google}. This unsolved challenge is technically very challenging to address---and potentially impossible to avoid---but increasing transparency in search results is key for progress.

The above-described open-ended and closed-ended mechanisms are orthogonal to the level of personalization, which includes the following exposure mechanisms.

\subsection{Content availability (generic exposure)}
\label{subsec:generic-exposure}
Back in the day, there was a standard view for everyone navigating through a web-page; just like today, everyone would see the same Wikipedia page if looking for a specific term in a specific physical region and language~\citep{Wikipedia}. Being exposed to generic content brings benefits such as preventing a biased machine from deciding what children should be exposed to; however, it still carries an overwhelming amount of information online. Search engines and personalized recommendations tackle some of these challenges by learning about the user and understanding what they may enjoy better~\citep{Netflix-help, Netflix-personalization}.

\subsection{Content recommendation (targeted exposure)}
\label{subsec:targeted-exposure}
To optimize for a ``better'' online experience, recommending systems have become the primary tool for ``improving'' this experience. Personalizing learning recommendations and EdTech have become more prominent during the last few years~\citep{deschenes2020recommender}. One of the main benefits of personalized learning is the scalability it can achieve. On the flip side, these algorithms may be locking children in stereotypes from a very early age, stereotypes that may be temporary proper or mistaken, and which may be impacting children's life opportunities (by reducing the exposure to only what aligns with their math-based profile) and social mobility as a consequence of reduced opportunities~\citep{alg-injustice-in-ed, eubanks2014want, o2016weapons}.

By profiling children and routing them to specific experiences, algorithms may also reduce their agency for self-exploration and development~\citep{barassi2020child}. Children are in a crucial development stage where they also want to be socially accepted and find their space and uniqueness in society. These potential effects of interacting with A.I. bring ethical concerns for how much are models allow children to freely choose the values they want to be associated with (moral autonomy), as well as how much can they control their narrative in different contexts (contextual integrity) as opposed to models directing content ranking according to what may be more profitable for the platform~\citep{livingstone2018children, barassi2019datafied}.

\subsection{Content moderation (controlled exposure)}
\label{subsec:restricted-exposure}
As sometimes parental control tools allow to~\citep{livingstone2017maximizing}, systems can be trained to control (through restricting or enabling) exposure to content. Thanks to moderation tools, we see big tech companies addressing the spread of fake news or hate speech. We also see systems such as Youtube kids or search engines designed for children that are developed to be safe for children of multiple ages, allowing large-scale information access for youth---in part allowing other beneficial types of exposure. Nevertheless, there is an assumption that the service providers know what safety means for each community, how risks are presented in their platforms, and what risks are more likely to happen within specific communities.

In their latest work,~\citet{haimson2021disproportionate} show that content moderation is not equitable for marginalized social media users. For example, they show that platforms---like Facebook, Instagram, and Twitter---bias their content moderation against transgender and Black users by removing their posts involving content related to expressing their marginalized identities despite following site policies.

Further, content moderation is not necessarily only about restricting content for children. In other words, content moderation does not mean delegating all the decision-making power to a platform; it can also be designed to allow users to have agency and control of their experience~\citep{bhargava2019gobo, saldias2019tweet, lai2022human}.

\textbf{Community-values-guided moderation}. From a machine-learning perspective, training a model for classification tasks (e.g.,  abinary decision for content appropriateness) rely on a clear objective or common sense (e.g.,  differentiating a dog from a cat; or a negative movie review from a positive movie review). However, there are other similar tasks where biases influence decisions that can cause direct harm to humans or when the ground truth / gold standard depends on the specific intended audience.
    
Content moderation is, arguably, an instance of such a task, where content appropriateness can depend on the target community and their values. In this scenario, each family (or community) should be empowered to define different rules to explain why a piece of content should be filtered (out or in) in their children's news feeds, instead of having a universal (top-down) model that rules what is right from wrong independent of the communities~\citep{milan2019big}, as these universal approaches are proven to make mistakes that can lead to even amplification of hate~\citep{underresourcedmod, milan2019big}.

Here we propose to frame moderation as a \textit{highly human-controlled and interpretable} tool, where families have agency over and visibility of the specific content their children consume online.\\

Whenever we browse the internet, we may encounter multiple combinations of the types of exposure to content described above. For example, an Instagram news feed shows personalized content (targeted exposure---section~\ref{subsec:targeted-exposure}) through a stream of content (that many believe to be open-ended exposure---section~\ref{subsec:open-ended-exposure}---while in reality it is optimized for engagement). When looking for a specific hashtag (\#) that can become a user-narrowed search (closed-ended exposure---section~\ref{subsec:closed-ended-exposure}), where content is not much moderated other than following Instagram universal moderation rules~\citep{milan2019big}. From our understanding of the content-exposure landscape and ethical A.I., we envision a world where many worlds fit. Not only those worlds dictated by social-media platforms, which have the power to shape children---most times---disregarding the responsibility to allow for their self-determination~\citep{costanza2018design}.

\section{Designing Content Moderation Models With and For Children}
\label{sec:design-proposal}

In this section, we illustrate opportunities to address core ethical considerations discussed above, along with design considerations for tools that may be deployed within systems that intentionally or unintentionally interact with or affect children. Without much loss of generality, the specific scenario of analysis is valued-guided moderation of text-based content designed for and with children and families, which we call child-centered content moderation. By the end of this section, we delineate opportunities to intend this system for and with children.

\subsection{Child-centered control: enabling and moderating content}
Content moderation is one of those tasks where biases influence decisions that can cause direct harm to humans (e.g., by discouraging, invalidating, or silencing people's experiences and opinions shared online) or when the standard depends on the specific target audience (e.g., language that may seem rude for one community may be perceived as adequate in another one). We argue that, when content appropriateness depends on the audience community (e.g., children age-groups or family-level communities).

\textbf{Our driving design question is}: ``How can we help facilitate a child-centered content curation process to (1) include aspects that families (or communities) value for their children and (2) provide interpretable rationales about why a piece of content may be appropriate or inappropriate for them, so that we can scale (according to systems and human needs) the moderation process by being assisted by A.I.?''.

One approach could be to create a classifier to tell us how appropriate a piece of content is. In fact, there are good models already out there that help detect hate speech, toxicity, profanity or violence~\citep{perspective-api}. These models, as we have seen in previous work, extract textual cues and meaning to determine a class (a.k.a., violation type). However, these have a nearsighted view of what moderation and control imply, namely only filtering out toxic or violent content---taking a top-down approach where powerful companies have decided what is to be filtered out. Here we re-frame this challenge as an opportunity. What about allowing families to guide content-moderation strategies according to what they value? (as opposed to agreeing with parental control models that may unnecessarily limit children's exposure to content by activating what companies think is child-proof or age-appropriate.)

Further, what about pursuing a healthy media diet~\citep{jackson2019you} by setting the proportion of values we want our children to be exposed to? E.g., balancing between content depicting drinking, drugs \& smoking, or consumerism (potentially aiming to reduce it) and content with positive characteristics, such as positive educational value, positive messages and positive role models (potentially aiming to increase it)? Can we empower families and caregivers with this level of granularity? We argue yes, and substantiated by this research and development companies like~\citet{commonsensemedia}, in the next section present an approach for this.

\subsection{Prompting value scenario}

Why are companies in charge of moderating most of the content children and adults consume online? We argue that each family or caregiver may want to allow their children to different levels of content characteristics (include more of some or less of others). Further, different families and communities may perceive differences in positive messages, role models, or consumerism. Even more so, value definition can evolve within and across families.

To ground the design of a child-centered moderation tool---and to emphasize implications for stakeholders---we follow a value-sensitive-design research strategy~\citep{friedman2017survey}, through imagining the following value scenario:

\begin{shadedquotation}
\begin{quote}
Laura (girl, 14 years old) uses her smartphone to access social media. By default, she sees a stream of content that is also targeted to retain her attention and increase her time spent on the platform.\\

Laura's parents follow her social media diet through a parental-control app that allows them to control and moderate the different content Laura is exposed to. Thanks to a values-aware moderation tool, Laura's parents realize her news feed is skewing towards drinking and consumerism. They want to know from what stance drinking and consumerism are being displayed. Is it preventing drinking habits in youth? Is it challenging consumerism? or is it encouraging these habits?\\

Laura's parents have not been trained as computer engineers. Hence, they do not have the tools to answer this question if not for (1) manually going through all the content Laura is exposed to or (2) having intelligible access to this feature in their parent-moderation system view.\\

Further, as Laura is getting into high school and close to graduating, her community---school and parents--- wants to expose her to diverse skills and potential role models that reflect some of her interests and other skills she may be interested in acquiring; using a family- / community-guided moderation approach.\\

As designers of this moderation tool, we realize the risk of parents forcing their children to specific pathways, which is why these family-guided recommendations influence the content appearing on Laura's news feed only to some extent, allowing accidental exposure to opportunities their families may not think for them.    
\end{quote}
\end{shadedquotation}

\subsection{Design considerations}
\label{subsec:design-considerations}

For this values-aware moderation tool to be empowering for families, it needs to (1) respect each family's values and (2) provide intelligible rationales about the process and moderated content. This human-in-the-loop scenario raises a tension between large-scale automatic moderation and rapid and fresh access to content. While an end-to-end machine learning system that assumes all children and families align with the same values and priorities is not real and/or desirable, we argue that we can still automate parts of the moderation process without reducing decision makers' agency (e.g., caregivers' and children's), by creating assistive tools intended to deepen the understanding and agency of children as digital citizens~\citep{third2016rethinking}.

To respond to these considerations, we propose a system design targeted to foster self-identity and community development and counteract oppression on three levels~\citep{costanza2018design}:\\

\begin{enumerate}
    \item \textbf{Personal biography}: this moderation tool should not deny a child's identity by inexplicably forcing content to them that are unaligned with those of their family or children themselves.\\
    
    \item \textbf{Community and cultural context}: this moderation tool should prioritize those values diverse families care about, as our goal is to work with them to create a better online space for children instead of setting a universal moderation agenda that foster certain kinds of communities while suppressing others~\citep{bhargava2019gobo}.\\
\end{enumerate}

Rapid prototyping, formative evaluation, and field testing with families and children can be an effective means to detect whether this novel values-guided approach to moderation systems is allowing them to control the content aspects they wish for their children and to surface and evaluate unintentional biases throughout the design process~\citep{feuz2011personal}.

Specific design considerations that will help address the challenges presented above are as follows.\\

\subsubsection{Access to diverse perspectives}

Western-centric values and social understanding can fail to recognize non-mainstream ways of knowing and understanding the world through individual life experiences. This brings a large-scale concern as the most prominent content platforms worldwide are based in the USA (Google, Facebook, Instagram, Twitter, among others). By deploying their top-down universal moderation values and policies, we remain in a blind spot that can be addressed as we delegate decision-making power to communities and families themselves. Even assuming that our system is successfully implemented in the USA, theory tends to travel badly, and more often than not, we can fail to acknowledge the specificities of distinct geographies, cultures, communities, and families~\citep{milan2019big}.

While not easy to address, as we are thinking of building on top of these Western-centric media platforms, we have opportunities for including diverse perspectives and values in the moderation schema. First of all, by allowing families to control the level of each content type they wish for their children, we have the capacity to help them portray their own values in their children's content consumption. Secondly, prototyping and continuously evaluating this system by gathering feedback from various families can strengthen the inclusion of diverse perspectives.\\

\subsubsection{Accountability through transparency}

To pursue accountability, we need more than ethics, whose mission is not to regulate but establish ethical principles~\citep{resseguier2020ai}. We intend for our value-sensitive moderation tool to pursue accountability through transparency as it empowers external agents to audit and challenge this system's internal functioning. Yet, this is not enough. We still need clear regulations that guide us to empower families to advocate for their rights and responsibilities in our system. This additional structure will allow those affected to trust in the accountability system and neither be nor feel submissive to our or any A.I. systems.

Major risks that we face in light of accountability include 1) being able to recognize when our system is not achieving its purpose (section~\ref{sec:design-proposal}) and act upon that, 2) standing by our design principles that generate trust in our intended audience (a.k.a., families) (section~\ref{subsec:design-considerations}), and 3) working in our mission of increasing transparency and not hiding design pitfalls.

Addressing sources of unintentional harm and discrimination and remedying the corresponding deficiencies will be difficult technically, difficult legally, and difficult politically. Yet, there is a lot that we can do internally, like including a culture of internal and external audit and evaluation of our systems---technically and in the ways they affect children~\citep{nissenbaum1996accountability, resseguier2020ai, shneiderman2020human}.\\

\subsubsection{Social implications of design}

The proposed content moderation system is meant to be a platform to facilitate family-guided moderation for their children to improve their safety and experience online. This system will allow families to request their children's data to be removed from the system at any point. While that may reduce the effectiveness of our algorithms when recommending or removing content to their children's news feed, we believe in people making their own choices when it comes to interacting with recommendation and moderation tools---in the end, their decisions are affecting their family.

Looking into economics and social implications of design, \cite{eubanks2014want} raises two critical questions on digital human rights to reflect upon before we push forward with our systems, adapted to this scenario (namely, replacing ``the poor'' for ``children and families''):
\begin{enumerate}
    \item Does the tool increase the self-determination and agency of children and families?
    \item Would the tool be tolerated if it was targeted at non-poor children and families?
\end{enumerate}

We argue that, as presented here, the proposed child-centered, family-guided moderation system is a decisive step forward in addressing both these questions.

First of all, our system would be designed---and more importantly evaluated--deliberately to increase families' agency on their children's content. Further, it increases transparency and allows children to understand why some content appears and other is hidden. Also, it will enable children to guide, through their families, the content characteristics they want to consume more or less frequently, increasing self-determination by (children) influencing and (parents) controlling their news feeds.

Secondly, this moderation system is not targeted at a specific group of families but rather aims to learn what different families and communities care about and add these values as filters they could control for. We acknowledge that the initial set of values we set for the system will (define a matter of course) represent culturally-biased decisions. To address this, the design process can aim at reducing these biases by running prototypes with different communities before deploying it to them. As more and more families adopt this system, designers and engineers will be better equipped to understand and serve their needs. It is essential to highlight that this system is not imposing a set of values that people should consider but rather allowing families to control for a list of values that the system allows. That list is intended to grow as the system grows.

A big challenge that needs addressing is understanding the risks associated with collecting data from children, data that could be later---against to core design values---used for profiling them further into their lives~\citep{barassi2019datafied}. Concretely, as children's identity is continuously developing---and now recorded through their social media behavior or their parents' posting profiles---these data traces can be used (and are being used)~\citep{fb-2016-bbox, uk-debacle-2020} to categorize children and predict future behavior and performance potential in their daily lives. For example, as part of school or job applications or defining their insurance policies. Therefore, as a baseline measure towards mitigating this risk, deployment requires accountability towards not sourcing or joining any children-data marketplace and having internal policies about anonymization to prevent any children from being identified.\\

\subsubsection{With and For children}

Our guiding principles acknowledge that families have rights over their children's online experiences. But unfortunately, the amount of control---that parental tools offer today--- fails to recognize that these families and their communities may intend diverse and distinct sets of values for their children.

We propose to pursue fairness through awareness and transparency as we aspire to provide families with tools to act upon content moderation challenges---as opposed to hide them and wish we internally (and hiddenly) produce the best results~\citep{dwork2012fairness}. Further, ignoring the fact that children are shapeable and developing their identity fails to acknowledge their citizenship status on social life and media~\citep{barassi2019datafied}.

Fundamentally, evaluation methods for proposed moderation algorithms and systems need to acknowledge children on their role as humans with values, beliefs, needs, and as full digital citizens. This is reflected in that the rapid prototyping process---mentioned earlier in this section---calls to include feedback rounds with children and families not as a target group for our envisioned system but rather as rightful digital citizens.\\

\subsubsection*{With Children}

Children are the ultimate stakeholders of this system. They may be aware of their experiences beyond what they can express in words and are experts at their lives~\citep{zaman2020designing}. Predominantly, in systems for children, the decision-making power resides entirely in adults who develop, design, and regulate technologies for young people. However, we mean ``with'' children as it is not only designers but also children along with their families who ought to have faculty to control the key features of this content moderation system and its design process. Increasing children's decision-making power throughout the design process has the beneficial potential to help mitigate treating children as ``others'' for whom the product is designed as mere subjects in need of help. Aligning~\citet{zaman2020designing}'s youth-centered design opportunities and risks with~\citet{friedman2017survey}'s value-sensitive design approach is a promising path ahead for working with children in a more fruitful content moderation framework.

Furthermore, we account for research with children and parents, which shows that age-based regulatory approaches that seek to protect children's data via an age threshold prove effective primarily among young children compared to teens and older children~\citep{livingstone2018children}.\\

\subsubsection*{Privacy by design}

Involving children and their families in privacy by design endeavors helps surface risks and harms to be detected before these become emergent biases and unnecessary risks~\citep{feuz2011personal}. Furthermore, inviting children to learn and influence the prototyping and downstream moderation processes can benefit them in the long term. By empowering children and meaningfully increasing their agency in the design and decision-making process, they can acquire new competencies, including knowledge, skills, and critical and constructive attitudes toward emerging technologies~\citep{zaman2020designing}. A simple rationale behind this is that children and their families must live with the effects of the proposed system; hence, we argue that they should have the right to control their usage and guide design evaluation. Further, in the future, designing to protect the privacy and rights of all users may work better than trying to identify children among users so as to treat them differently privacy-wise~\citep{livingstone2018children}.

Note that we need to beware of possible unintended side effects of adult designers working with children. Adults are still responsible for the decisions, and we cannot fail to believe that including children relieves adults' responsibility by delegating it to children (as per their input).\\

\subsubsection*{For Children}
Creating a value-guided moderation system ``for'' children opens doors for adults to increase control of their own feeds. It can prove more efficient and effective to develop systems for more vulnerable communities (i.e., children) and then focus on addressing adult's needs (as opposed to addressing adult's needs first and then tackling issues with children as an afterthought)~\citep{livingstone2018children}.

We acknowledge that, just as in GDPR~\citep{livingstone2018children}, the proposed moderation system relies on families with conscientious parents and dutiful children. However, the messy world of real families---who may lack time, share devices with one another, or have internal conflicts--- fails to fit the engagement need for family-guided moderation. Delegating responsibility to schools, chosen by caregivers to form their children, can alleviate the necessity of caregivers' time to set up and keep track of a system like this for their families (reason why we talk both about families and communities as leading this effort).

At the same time, while GDPR acknowledges that children's data is worthy of protection, it still fails to address unresolved challenges, like:
\begin{enumerate}
    \item Children's media literacy: how aware are children of the risks and rights of processing their personal data? Are parents aware?
    \item Commercial profiling: there is a lack of regulation relating to children's data. This implies dealing with stakeholders and grappling with technical capabilities for safe and guaranteed anonymization.
    \item Nature of family relations: how ready are families to act according to current regulations? How literate are they in privacy and their responsibility?
\end{enumerate}

Under these challenges, assuming that many parents and caregivers may not be able to dedicate enough time to control and regulate their children's online content diet, the proposed moderation tool should offer a default set of settings that comply with the law of the communities in which this is deployed. As explored by~\citet{livingstone2017maximizing}, a moderation tool that enables more content does require more skilled parents who are aware of the online opportunities available and recognize how to activate them and control their risks.\\

\section{Conclusion}
\label{sec:conclusion}

In this work, we describe a new system aimed at facilitating child-centered, family-guided moderation through specific design considerations, specifically focused on children's personal biography, community and cultural context, systemic and social institutions, access to diverse perspectives, transparency through interactivity and human control, accountability through transparency, social implications of designing this system, and design and privacy considerations to design with and for children and families. To prompt these specific considerations, we use a value scenario to help highlight and reflect on the main stakeholders and how this system can be of help or cause harm. The presented value scenario and reflection on it offer researchers interested in the intersection of A.I. and child-development or child-safety systems an anchor to help ground and reveal challenges and opportunities for their work.

While designing for good can mean different things for different people, the proposed system is intended for good because it provides a baseline platform to increase families' and children's agency and control over the content children consume or are exposed to on the internet. Through this work, we surface critical ethical considerations, challenges, and opportunities for the presented system. We argue that this child-centered system works towards increasing socially preferable outcomes, allowing families to decide what values they wish for their children, instead of having a universal top-down approach imposed by moderation and recommendation algorithms deployed by big tech/media companies.

Implementing a successful child-centered, family-guided moderation system will require continuous prototyping and families' participation to help align its socio-technical development to the specific design considerations (opportunities and challenges) we outline in this work. In addition, continuous evaluation should enable us to learn from the systems' and development's successes and opportunities to serve better those families and communities intending to use a system like this one.

We hope this work serves as an example and stepping-stone for designers and researchers who undertake the mission to increase children's safety online by focusing on the challenges of youth interacting with intelligent agents that are currently controlling the content they get exposed to, and by extension, the values and opportunities they have access to---and, in turn, contribute in unimposing yet crucial ways to the fight for children online safety from an A.I. perspective.

\bibliographystyle{apacite}
\bibliography{interactapasample}
\end{document}